\documentclass[11pt,a4paper]{article}
\pdfoutput=1
\usepackage{jheppub}

\usepackage[utf8]{inputenc}
\usepackage{hyperref}
\usepackage{amsmath}
\usepackage{amsfonts}
\usepackage{amssymb}
\usepackage{graphicx}
\usepackage{color}
\usepackage{verbatim}
\usepackage[normalem]{ulem}

\def\t{\ev}

\def\ev{\theta}

\def\A{A}
\def\H{M}
\def\S{H_0}
\def\M{H}

\def\De{S}

\def\Pim{p}

\def\x{\kappa}
\def\y{\eta}

\def\sp{S}

\def\Dov{D_{{\sf ov}}}
\def\DW{D_{{\sf W}}}

\def\eq#1{Eq.~(\ref{#1})}
\def\fig#1{Fig.~\ref{#1}}

\def\sutwo{SU(2)}

\DeclareMathOperator{\re}{\mathfrak Re}
\DeclareMathOperator{\im}{\mathfrak Im}

\definecolor{orange}{rgb}{1.,0.65,0}
\definecolor{green}{rgb}{0,.7,0}

\newcommand{\Dw}{\DW}
\newcommand{\Gl}{Gl}
\DeclareMathOperator{\sgn}{sign}
\DeclareMathOperator{\diag}{diag}
\DeclareMathOperator{\erf}{erf}
\DeclareMathOperator{\erfc}{erfc}

\newcommand{\sect}{Sec.~}

\begin{document}

\title{
Level spacings for weakly asymmetric real random matrices and
application to two-color QCD with chemical potential
}

\author{Jacques Bloch, Falk Bruckmann, Nils Meyer, Sebastian Schierenberg}
\affiliation{Institute for Theoretical Physics, University of Regensburg, 93040 Regensburg, Germany}

\subheader{\hfill\normalsize July 19, 2012}

\abstract{
We consider antisymmetric perturbations of real symmetric matrices in the context of random matrix theory and two-color quantum chromodynamics. We investigate the level spacing distributions of eigenvalues that remain real or become complex conjugate pairs under the perturbation. We work out analytical surmises from small matrices and show that they describe the level spacings of large random matrices. As expected from symmetry arguments, these level spacings also apply to the overlap Dirac operator for two-color QCD with chemical potential.}

\maketitle
\flushbottom

\section{Introduction}

Random matrix theory (RMT) has a remarkably wide range of applications in mathematics, physics and beyond. It describes universal statistical quantities that are not determined by the specific dynamics of a given system, but rather by its anti-unitary symmetries. One of the most prominent RMT quantities is the level spacing distribution $P(s)$, the probability of two neighboring eigenvalues to appear at a distance $s$, which is universal after unfolding (to refer to an eigenvalue density normalized to one). For the classical Gaussian RMT ensembles, i.e.\ hermitian matrices with Gaussian probability distribution of the elements, the spacing distributions can be evaluated for any matrix size \cite{Mehta:2004}. The results are very close to the level spacings of small matrices, a fact which is known as Wigner's surmise \cite[p.~199]{porter:1965}. The spacing distributions $P(s)$ of the smallest matrices are linear/quadratic/quartic in $s$ times an exponential decay in $s^2$ (see e.g.\ \eq{eq:surmiseon1}), where the power in the repulsion for small spacings -- which equals the Dyson index $\beta\in\{1,2,4\}$ of these ensembles -- counts the degrees of freedom per matrix entry in the ensemble. 

In particular, RMT applies to quantum chromodynamics (QCD), where it describes the spectrum of the Dirac operator; for a review see \cite{Verbaarschot:2000dy}. The crucial feature is that QCD (with massless quarks) obeys a global chiral symmetry, which is spontaneously broken. In the $\varepsilon$-regime, a finite volume regime dominated by the corresponding Goldstone bosons (pions), the equivalence with RMT has been proven in Refs.~\cite{Osborn:1998qb,Toublan:1999hx,Basile:2007ki}. The appropriate RMT describing the microscopic spectral properties of QCD in this regime is the chiral ensemble. For bulk correlations, i.e. away from the origin, the chiral symmetry is no longer important \cite{Fox:1964, Nagao:1991, Nagao:1992} and universal properties of QCD spectra, such as level spacings, can be described by the standard Gaussian RMT ensembles \cite{Halasz:1995vd,Pullirsch:1998ke}.

While the conventional RMT ensembles describe chaotic systems with definite symmetries, physical systems are often `not so ideal'.  We briefly describe cases where the symmetry of the system deviates from this ideal situation. When no analytical expressions are available for the correlations in such transitions, one can have recourse to surmises.

Firstly, physical systems may obey a certain \textit{anti-unitary symmetry} only approximately or contain several parts with different anti-unitary symmetries. The latter leads to transitions from one symmetry class to another.  Including the Poissonian ensemble, which is the equivalent for generic integrable systems, the corresponding spacing distributions have been shown to fulfill a generalized Wigner surmise \cite{Lenz:1991, Schierenberg:2012ut}. Such transitions were observed at high temperatures in two- and three-color QCD \cite{Garcia-Garcia:2006gr, Kovacs:2009zj, Kovacs:2010wx, Bruckmann:2011cc} and are expected in the continuum limit of the staggered lattice Dirac operator in two-color and adjoint QCD \cite{Follana:2006zz, Bruckmann:2008xr}. We will show a mixed-symmetry surmise at work for two-color QCD at imaginary chemical potential.

Secondly, \textit{hermiticity} can be broken in physical systems, as is the case for QCD with real chemical potential. In this case the determinant of the Dirac operator becomes complex, which hampers progress in numerical simulations of this system. Non-hermitian Gaussian random matrices were first discussed by Ginibre \cite{Ginibre:1965zz}. For the associated complex eigenvalues one can define a nearest neighbor individually and study the corresponding spacing distributions. For the complex Ginibre ensemble the level spacings do not fulfill a surmise from $2 \times 2$ matrices (even though they are always cubic for small $s$) \cite{Grobe:1988,Akemann:2009my}. The spacings of this ensemble were compared to QCD spectra and showed agreement only for a small range of the chemical potential \cite{Markum:1999yr}.

In QCD the chemical potential drives the breaking of the anti-hermiticity. An adequate random matrix model describing such a situation consists of two matrices where the breaking of the hermiticity is governed by a real  parameter \cite{Stephanov:1996ki,Osborn:2004rf}.  
The eigenvalue densities and lowest eigenvalue distributions of such ensembles have been worked out in the microscopic, {\it weakly} non-hermitian limit (where the microscopic eigenvalues $z N$ and perturbation $\mu^2 N$ are kept fixed as the matrix size $N \to \infty$) \cite{Splittorff:2003cu,Osborn:2004rf,Akemann:2004dr,Akemann:2007yj} and successfully compared to quenched QCD data \cite{Bloch:2006cd,Akemann:2007yj}.

The treatment of real random matrices is known to be an especially arduous task. The joint probability function of the eigenvalues for the real asymmetric ensemble (also called the real elliptic Ginibre ensemble) was calculated in \cite{Lehmann:1991} and generalized to the chiral case in \cite{Akemann:2009fc}.  Only recently, the correlation functions were worked out for the real Ginibre ensemble \cite{Forrester:2007,Sommers:2008,Borodin:2009}, the real elliptic Ginibre ensemble \cite{Forrester:2008} and the real chiral ensemble \cite{Akemann:2009fc,Akemann:2010mt,Akemann:2010tv}. 

In this work we investigate the use of surmises to describe the level spacings of such real random matrix ensembles. More specifically, we examine the case where the unperturbed matrices are real and symmetric, i.e., they belong to the Gaussian orthogonal ensemble (GOE, $\beta=1$), and the perturbations are real and antisymmetric. The perturbations are assumed to be small, such that the imaginary part of typical eigenvalues is at most of the order of the mean separation between neighboring eigenvalues along the real axis (whereas in the Ginibre ensemble hermiticity is maximally broken). For real matrices the flow of eigenvalues is somewhat special, because two real eigenvalues need to coalesce to `give birth' to a complex conjugate pair, see below (this is not so for the complex (GUE, $\beta=2$) or symplectic (GSE, $\beta=4$) ensembles: in the former the eigenvalues are not restricted to form complex conjugate pairs, but can be distributed arbitrarily in the complex plane; in the latter the unperturbed matrices have two-fold degenerate eigenvalues due to Kramers' degeneracy and these split up and become complex for arbitrary small anti-hermitian perturbation). Actually, the original GOE eigenvalues are \textit{attracted} by an antisymmetric perturbation. For eigenvalues that meet in such a process and become a complex conjugate pair, a spacing can be uniquely defined as the distance between them.  As we will show, the distribution of these spacings for large random matrices can be approximated very well by that of small matrices. The complex eigenvalues leave a trace in the statistics of the real eigenvalues, too. We will work out surmises for the spacings among the latter and show that they also apply to large matrices. We interpret these findings in the same way as for other Wigner surmises, namely  that the dynamics of a neighboring pair is not much influenced by the effect of all other eigenvalues.

As an application we analyze spectra of two-color QCD, i.e., with gauge group \sutwo. As the latter is pseudo-real it ensues that the massless continuum Dirac operator for fermions in the fundamental representation belongs to the GOE universality class (see Sec.~\ref{subsect:symm_and_herm}).
This property holds for vanishing chemical potential. Although the chemical potential perturbs the anti-hermiticity of the Dirac operator, it keeps its anti-unitary symmetry. Therefore, the random matrix results discussed above apply to two-color QCD at finite chemical potential, both in the continuum and on the lattice with the overlap operator.

This paper is organized as follows. After describing the symmetries and corresponding spectra of matrices in the next section, \sect\ref{sect:perturbation_theory} is devoted to the eigenvalue dynamics in perturbation theory. In Sec.~\ref{sect:surmises} we work out the surmises for the different types of spacings. These are then matched to data from large random matrices with mixed hermiticity in \sect\ref{sect:rmt} and to lattice two-color QCD data with chemical potential in \sect\ref{sect:qcd}. We also measured spacings at imaginary chemical potentials and discuss the comparison to a mixed-symmetry surmise in the last subsection of \sect\ref{sect:qcd}.  We summarize our findings and conclude in \sect\ref{sect:summary}.

\section{The setting: symmetries and spectra of RMT ensembles}

Our unperturbed random matrices $\S$ shall be real and symmetric. We therefore choose them from the GOE, so that the entries are Gaussian random numbers with mean zero and variance
\begin{align}
 \langle (\S)_{ii}^2 \rangle = 1\quad\mbox{and}\quad \langle (\S)_{ij}^2 \rangle =
\frac12
\quad (i\neq j)\,.
\end{align}

Gaussian random matrix ensembles can be characterized by whether they commute  with an anti-unitary $V=UK$, where $U$ is unitary and $K$ is the complex conjugation operator
\begin{equation}
 [\S,V]=0\qquad \mbox{equiv. to} \qquad \S^\ast = U^\dagger \S U\,,
 \label{eqn symm}
\end{equation}
and whether $V$ squares to $1$ or $-1$. For the GOE one has
\begin{equation}
 V^2=1 \qquad \mbox{equiv. to} \qquad U U^\ast=1\,,
 \label{eqn goe}
\end{equation}
from which it follows that the matrices $\S$ can be chosen real.

\begin{figure}[t]
\centering
  \includegraphics[width=0.8\linewidth]{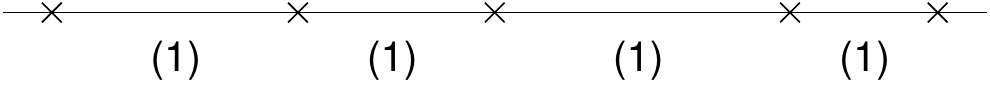}
  \vskip1cm
  \includegraphics[width=0.8\linewidth]{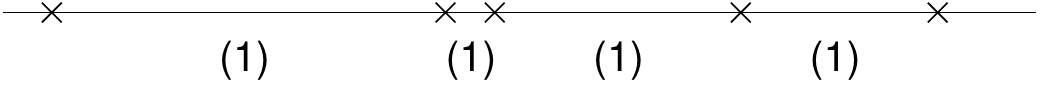}
  \vskip0.6cm
  \includegraphics[width=0.8\linewidth]{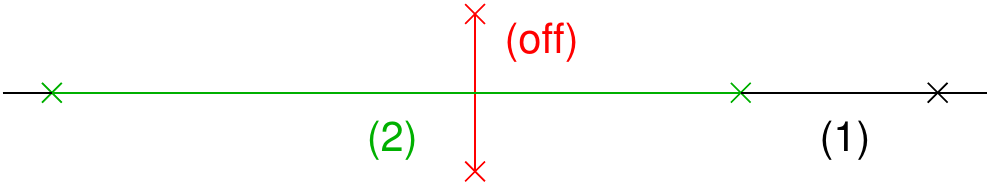}
 \caption{Evolution of eigenvalues of a typical random matrix with increasing non-hermitian part (schematically). The top panel depicts a part of the spectrum for the unperturbed case, i.e., for a real symmetric (GOE) matrix. Towards the bottom panel the antisymmetric part increases, which results in some eigenvalues getting very close and finally becoming complex (the rightmost eigenvalues are spectators).
 Different types of level spacings are sketched: (1) spacings between original eigenvalues and those that remain on-axis and have no interspaced complex conjugate pair, named $P^1_{\textrm{on}}$ and described by the pure GOE surmise, (2) spacings between on-axis eigenvalues with interspaced complex conjugate pairs, named $P^2_{\textrm{on}}$, (off) spacings between complex conjugate pairs called off-axis, named  $P_{\textrm{off}}$. Note that for QCD spectra the real and imaginary axes are swapped.}
 \label{fig general spectra}
\end{figure}

In this paper we will study random matrices from the GOE undergoing a real antisymmetric perturbation with a tunable coupling parameter. The characteristic equation of such a mixed matrix has real coefficients and consequently the eigenvalues are either real or come in complex conjugate pairs. Typical spectra of such matrices are depicted schematically in \fig{fig general spectra} together with different types of spacings between them. In the following, the real eigenvalues will be called `on-axis' and those with a non-zero imaginary part `off-axis'.  This notation also applies to the case of QCD, where due to the anti-hermiticity of the Dirac operator at vanishing chemical potential the axes are swapped (then on-axis refers to purely imaginary eigenvalues and off-axis to those evolving from there into the complex plane). Obviously the spacings between on-axis eigenvalues and off-axis eigenvalues will be treated separately. Among the former we will further distinguish between those with and without an interspaced complex conjugate eigenvalue pair. This distinction only makes sense if the hermiticity of the initial GOE matrix is only mildly broken (since otherwise the off-axis eigenvalues move arbitrarily in the complex plane and cannot be assigned to a single on-axis pair anymore), and we will therefore only consider this case here. This is also why we do not consider the rare case of two or more interspaced eigenvalue pairs. 

All our level spacing distributions are subject to the normalizations $\int_0^\infty ds \, P(s)=1$ and $\langle s \rangle = \int_0^\infty ds \, s \, P(s)=1$.

\section{Repulsion and attraction of eigenvalues in perturbation theory}
\label{sect:perturbation_theory}

To gain intuition about the dynamics of the eigenvalues, we consider a real symmetric matrix $H_0$, with real eigenvalues $\ev_i$ and corresponding real eigenvectors $|\psi_i\rangle$, perturbed by a real matrix $\lambda \H$ with $\lambda \in \mathbb{R}^+_0$. We compare the effect of symmetric and antisymmetric perturbations $\H$, whose matrix elements in the unperturbed basis,
\begin{align}
 \H_{ij}\equiv\langle\psi_i|\H|\psi_j\rangle\,,
 \label{Hij}
\end{align}
are real and symmetric or antisymmetric in $(i,j)$ too,
\begin{align}
 \H_{ji}= \pm\H_{ij}\,,\qquad
 \mbox{for }\H^T=\pm \H\,.
\label{eqn expvalues perturbation}
\end{align}
As is well known, ordinary perturbation theory up to second order yields for the eigenvalues,
\begin{align}
\ev_i \to \ev_i + \lambda \H_{ii} +
\lambda^2\sum_{j\neq i} \frac{\H_{ij}\H_{ji}}{\t_i-\t_j}\,.
\end{align}
The first order term induces a random walk of the eigenvalues without correlating them. Moreover, in the antisymmetric case such expectation values are zero, so there is no first order contribution.

When omitting the random walk term, the eigenvalue differences are
\begin{align}
(\ev_i-\ev_j) \to (\ev_i-\ev_j)\left[1\pm
2\lambda^2\frac{\H_{ij}^2}{(\ev_i-\ev_j)^2}+
\mathcal{O}\left(\frac1{(\ev_i-\ev_j)(\ev_{i,j}-\ev_{k\neq i,j})}\right)\right]\,.
\label{eqn second diff}
\end{align}
If $\theta_i$ and $\theta_j$ are much closer to each other than to any other eigenvalue, the third term in square brackets can be neglected in relation to the second one. Then, for real symmetric $\H$ (upper sign) the difference grows, which is the common knowledge that eigenvalues repel. 

In contrast, for real antisymmetric $\H$ (lower sign), which we will consider from now on, the eigenvalues attract. Thus in perturbation theory it is favored that eigenvalues coalesce.

We recall that our investigations so far are based on the eigenvalues $\ev_i$ of the unperturbed matrix, which are real and remain so as long as ordinary perturbation theory is valid.\footnote{Ordinary perturbation theory only yields real shifts of the eigenvalues to all orders.}

When two eigenvalues $\ev_i$ and $\ev_j$ are very close, one has to apply almost degenerate perturbation theory \cite{Gottfried:1966}.  Neglecting the second order contribution of all other eigenvalues, it gives for the perturbed eigenvalues
\begin{align}
 (\ev_i,\ev_j)
\to
\mbox{evs}\left[
 \left(\begin{array}{cc}\ev_i&0\\ 0&\ev_j\end{array}\right)+\lambda
  \left(\begin{array}{cc} 0 & \H_{ij}\\ -\H_{ij} & 0\end{array}\right)\right]\,,
\label{eqn deg pert}
\end{align}
where we have made use of \eq{eqn expvalues perturbation}, with $\H_{ij}$ still given by \eq{Hij}, 
and evs denotes the eigenvalues of the matrix. For the eigenvalue difference one obtains
\begin{align}
\De=\sqrt{(\ev_j-\ev_i)^2-4\lambda^2\H_{ij}^2}\,.
\label{eqn delta}
\end{align}
To second order in $\lambda$ the expansion of $S$ reproduces the result from second-order ordinary perturbation theory discussed above.

More interestingly, the argument of the square root can turn negative for some critical $\lambda$, above which the eigenvalue difference becomes purely imaginary. The latter is consistent with the fact that the complex eigenvalues of a real matrix come in complex conjugate pairs. Imaginary spacings occur for $\lambda$ bigger than
\begin{align}
 \lambda_\text{crit}=\frac{|\ev_j-\ev_i|}{2|\H_{ij}|}
\end{align}
(specific to each individual eigenvalue pair). In the exceptional case that $\H_{ij}$ exactly vanishes, this particular eigenvalue difference remains unperturbed for any $\lambda$ (in perturbation theory). For perturbations beyond $\lambda_\text{crit}$ the eigenvalue difference can be written as
\begin{align}
 \De= 2i|\H_{ij}|\sqrt{\lambda^2-\lambda^2_\text{crit}} = 2i|\H_{ij}|\sqrt{\delta\lambda(2\lambda_\text{crit}+\delta\lambda)}\,,\qquad \delta\lambda\equiv\lambda-\lambda_\text{crit}\,,
\label{eqn delta imag}
\end{align}
and increases with the excess $\delta\lambda$ of the coupling over the critical coupling. Therefore, the perturbation causes the eigenvalues to be repelled in the complex plane.

Moreover, the eigenvalues leave the real axis perpendicularly to it. This can be seen by an explicit calculation of $\ev_{i,j}$ in \eqref{eqn deg pert} for $\lambda>\lambda_\text{crit}$, but it also follows directly from the fact that the antisymmetric perturbation is traceless and thus does not change the center of mass of the eigenvalues. When considering antisymmetric perturbations of large matrices beyond perturbation theory, only the center of mass of all eigenvalues is fixed and a complex conjugate pair could in principle leave the axis at any angle, as its center of mass contribution could be compensated by the motion of the remaining eigenvalues. Our data confirmed that, in accordance with almost degenerate perturbation theory, the complex eigenvalue pairs move perpendicularly to the axis just after their creation. When the perturbation grows further, higher order effects set in and the complex conjugate eigenvalues can leave the trajectory perpendicular to the axis.

\section{Surmises}
\label{sect:surmises}

The idea of Wigner surmises is to approximate level spacing distributions of large random matrices by those of random matrices with smallest possible size, typically $2\times 2$ matrices. Note that in the perturbative treatment in the previous section we have focussed on a pair of nearest neighbor eigenvalues and thus ended up discussing $2\times 2$ matrices and the properties of their eigenvalues. This may be seen as an indication that surmises for mixed systems may apply to large random matrices.

We recall that we consider three different types of spacings in our (large random matrix and QCD) systems: on-axis spacings with one and without interspaced complex conjugate eigenvalue pair, and off-axis spacings (see \fig{fig general spectra}). For the first and third type we will deduce surmises from $2\times 2$ matrices, computing the corresponding spacings. The spacings of the second type will be computed from the joined probability density for $4\times 4$ matrices.

\subsection[Prelude: $2\times 2$ matrices and their spacings]{Prelude: $\mathbf{2\times 2}$ matrices and their spacings}

For size $2\times 2$ let us consider a real symmetric and traceless matrix perturbed by an antisymmetric one,
\begin{align}\label{eq:ortantisymm2x2}
 \M = \left(
 \begin{array}{ccr}
  -a & b \\ b & a
 \end{array}
 \right)+\lambda\left(
 \begin{array}{cc}
  0 & c \\ -c & 0
 \end{array}
 \right)\,,
\end{align}
with random numbers $a$, $b$ and $c$ that are Gaussian distributed with zero mean and unit variance. Note that the matrix $\M$ with $\lambda=0$ is equivalent to a $2\times 2$ GOE matrix, up to a common shift of the eigenvalues which does not alter the spacings. In the GOE the diagonal entries have twice the variance of the off-diagonal ones, but it can easily be shown that neglecting the part proportional to the identity, as we have done in $\M$ to simplify the computations, renders the variances equal. The eigenvalues of $\M$ are $\pm\sqrt{a^2+b^2-c^2\lambda^2}$, which, for a particular draw of random numbers, turn imaginary beyond a critical coupling
\begin{align}
 \lambda_\text{crit}=\frac{\sqrt{a^2+b^2}}{|c|}\,.
 \label{lambdacrit}
\end{align}
For $\lambda<\lambda_\text{crit}$ the spacing is $2\sqrt{a^2+b^2-c^2\lambda^2}$ and hence real, while for $\lambda>\lambda_\text{crit}$ the spacing is $2i\sqrt{c^2\lambda^2-a^2-b^2}$ and hence purely imaginary.

The probability that the matrix $\M$ has an imaginary spacing can be easily computed by integrating over all random matrix entries with the constraint that $\lambda$ is above its critical value,
\begin{align}\label{eq:imagspacprob}
 \Pim(\lambda) = \left(2\pi\right)^{-3/2}\int_{-\infty}^\infty\!\!da\, db\, dc \,e^{-(a^2+b^2+c^2)/2}\,
 \theta\left(c^2\lambda^2-a^2-b^2\right) = 1-\frac1{\sqrt{1+\lambda^2}}\,.
\end{align}
As expected, the limiting cases yield $\Pim(\lambda\to0) = 0$ and $\Pim(\lambda\to\infty)=1$.

\subsection{On-axis without interspaced complex eigenvalues}
\label{sect:onaxis1}
For on-axis spacings without interspaced complex eigenvalues our surmise is 
\begin{align}\label{eq:surmiseon1}
 P^1_{\textrm{on}}(s) =\frac{\pi}{2}\, s\, e^{-\frac{\pi}{4}s^2}\,,
\end{align}
which means that even in the presence of a (small) antisymmetric perturbation these spacings follow the GOE surmise governing the unperturbed spacings. This is so because \eq{eq:surmiseon1} is the distribution of \textit{real} spacings\footnote{The plural refers to the ensemble, as one representative of this random matrix has only one spacing, which will only be included here if real.} of $\M$ independently of the coupling parameter $\lambda$. Let us demonstrate the derivation of this distribution explicitly, as a typical example. We start with the distribution of the non-normalized spacing $\sp$
\begin{align}
 Q(\sp)&=\int_{-\infty}^\infty\!\!da\, db\, dc \,e^{-(a^2+b^2+c^2)/2}\,
 \delta\left(\sp-2\sqrt{a^2+b^2-c^2\lambda^2}\right)
 \theta\left(a^2+b^2-c^2\lambda^2\right)\nonumber\\
 &=2\pi\int_{-\infty}^\infty\!\! dc \int_0^\infty\!\! dr\,r\,e^{-(r^2+c^2)/2}\,
 \delta\left(\sp-2\sqrt{r^2-c^2\lambda^2}\right)
 \theta\left(r^2-c^2\lambda^2\right)\,.
 \label{eq:surmiseon_pre}
\end{align}
In the integration over $r$, the $\delta$-function replaces
\begin{align}
 r\to r_0= \sqrt{\sp^2/4+c^2\lambda^2}
\end{align}
with Jacobian
\begin{align}
\left|\left.\frac d{dr}\left(\sp-2\sqrt{r^2-c^2\lambda^2}\right)\right|_{r=r_0}\right|^{-1} =
\frac {\sp}{4\sqrt{\sp^2/4+c^2\lambda^2}}=\frac {\sp}{4r_0}\,.
\end{align}
Therefore, we obtain
\begin{align}
 Q(\sp) =\frac{\pi}{2} \int_{-\infty}^\infty dc \,\sp\,e^{-\left[\sp^2/4+c^2(1+\lambda^2)\right]/2}\,
 \theta\left(\sp^2/4\right)\,.
\end{align}
The integration over $c$ just gives a constant factor, and the $\theta$-function forces the spacing $\sp$ to be real as assumed. After normalization, we obtain $P^1_{\textrm{on}}$ from (\ref{eq:surmiseon1}).

\subsection{On-axis with interspaced complex eigenvalues}

\begin{figure}
 \centering
  \includegraphics[width=0.8\linewidth]{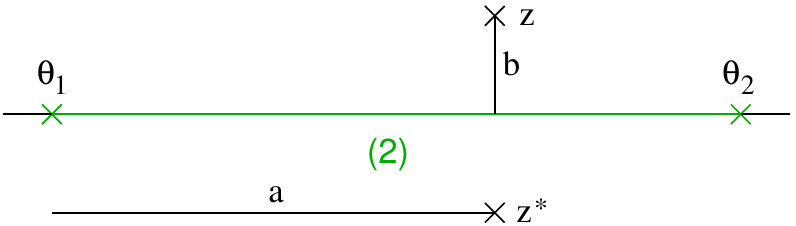}
  \caption{Spectrum of the $4\times4$ matrix used for the surmise for $P^2_{\textrm{on}}$.}
  \label{fig surmise spectra}
\end{figure}

When two on-axis eigenvalues have an interspaced complex eigenvalue pair, we expect the latter to have a noticeable influence on the spacing of the on-axis eigenvalues. To obtain a surmise for the distribution of these spacings, we consider a $4\times4$ matrix
\begin{align}\label{eq:ortantisymm4x4}
 \M = \S + \lambda\A\,,
\end{align}
with $\S$ taken from the GOE and an antisymmetric matrix $\A$ with probability density
\begin{align}\label{eq:probdistA}
w(\A) \sim e^{-\frac12 {\rm tr}\left(\A \A^T\right)}\,,
\end{align}
i.e. Gaussian distributed entries having the same variance as the off-diagonal entries of $\S$. This is the smallest possible matrix that can yield a surmise for this case as there are four eigenvalues which are relevant for the dynamics of the spacing.  The joint probability distribution of the eigenvalues of $\M$ depends on the number of real and complex eigenvalues.  For two real eigenvalues $\theta_1$ and $\theta_2$ and one complex conjugate pair $z$ and $z^*$, it reads up to a normalization \cite{Akemann:2010mt}
\begin{align}
 P(\theta_1, \theta_2, z, z^*) \propto e^{-\theta_1^2-\theta_2^2-z^2-(z^*)^2} \erfc\left(\frac{\sqrt{1+\lambda^2}}{\sqrt2\lambda}|z-z^*|\right)2i \Delta(\theta_1, \theta_2, z, z^*)\,,
\label{eqn jpd}
\end{align}
with the Vandermonde determinant
\begin{align}
 \Delta(\theta_1, \theta_2, z, z^*) = (z^*-z)(z^*-\theta_2)(z^*-\theta_1)(z-\theta_2)(z-\theta_1)(\theta_2-\theta_1)\,,
\end{align}
where it is assumed that $\theta_2>\theta_1$ and $\im\,z>0$.  To obtain the distribution of the spacing $\sp$ between $\theta_1$ and $\theta_2$, we set $\theta_2 = \theta_1 + S$ and introduce new variables $a = \re\,z - \theta_1\in[0,S]$ and $b = \im\,z\in[0,\infty)$, cf. \fig{fig surmise spectra}.
This results in
\begin{align}
 &P(\theta_1, \theta_1 + S, \theta_1+a+ib, \theta_1+a-ib)\\
 &\propto \sp\, b \left[\left(a-\sp\right)^2+b^2\right]\left(a^2+b^2\right) e^{-\theta_1^2-(\theta_1+\sp)^2-2(a+\theta_1)^2+2b^2}\, \erfc\left(\frac{\sqrt{2}\sqrt{1+\lambda^2}}{\lambda}\,b\right)\,.\nonumber
\end{align}
We integrate out $\theta_1$, $a$, $b$ and perform an irrelevant rescaling $S\to2S$ to obtain the non-normalized spacing distribution
\begin{align}
\int_0^\infty  db\int_{0}^{2\sp} & da \int_{-\infty}^\infty d\theta_1\, P(\theta_1, \theta_1 + 2 S, \theta_1+a+ib, \theta_1+a-ib) \notag\\
\propto  \sp e^{-3\sp^2} & \left\{ 
\sqrt{\pi} \, e^{\sp^2} \erf(\sp) \left[\lambda^2\sqrt{1+\lambda^2}\left(3\lambda^2+8\sp^2\right)+4 \left(\sqrt{1+\lambda^2}-1\right)\left(4\sp^4-8\sp^2+3\right)\right] 
\right. \notag
\\
& \left.  + 8\sp \left[\left(\sqrt{1+\lambda^2}-1\right)\left(2\sp^2-1\right) -\lambda^2 \sqrt{1+\lambda^2}\right] \right\}  \equiv Q(\sp; \lambda) \,.
\label{eq:surmiseon2_pre}
\end{align}
Properly normalized, the spacing distribution reads
\begin{align}\label{eq:surmiseon2}
 P^2_{\textrm{on}}(s;\lambda) = CD\,Q(Ds;\lambda)\,,
\end{align}
with normalization and scaling factors
\begin{align}
 C &= \left(\int\limits_0^{\infty}
d\sp\,Q(\sp;\lambda)\right)^{-1}=\frac{12\sqrt3}{\sqrt{\pi}}\left[\lambda^2 \sqrt{1+ \lambda^2} \left(9\lambda^2+8\right)+8 \left(\sqrt{1+\lambda^2}-1\right)\right]^{-1}\,,
\end{align}
and
\begin{align}
 D = C\,\int\limits_0^{\infty} d\sp\,\sp\,Q(\sp;\lambda) &= \frac{C}{72}\left[26 \left(\sqrt{1+\lambda^2}-1\right)-27\sqrt2\,\text{arccot}(\sqrt2)\right.\notag\\
&\left.+\sqrt{1+\lambda^2}\left(18\lambda^4+20\lambda^2+27\sqrt2 \left(1+\lambda^2\right)^2\text{arccot}(\sqrt2)\right)\right]\,.
\label{eq:surmiseon2_const}
\end{align}

In the limit $\lambda \to 0$, the spacing distribution reads
\begin{align}
 \hspace{-0.1cm}\lim_{\lambda\to 0} P^2_{\textrm{on}}(s;\lambda) = \frac{\x^2}{8\pi}\,s\, e^{-3\x^2 s^2}
\left[4\x^3s^3- 6\x s + \sqrt\pi\, e^{\x^2 s^2} (4\x^4 s^4-4\x^2 s^2 + 3)\,\erf(\x s)\right]\,,
\label{eq:surmiseon2_smallpert}
\end{align}
where
\begin{align}
 \x=\lim_{\lambda\to 0}D=\frac{22+45\sqrt{2}\,\mbox{arccot}\sqrt{2}}{16\sqrt{3\pi}} \approx 1.2453 \,.
\end{align}
The opposite limit $\lambda\to\infty$ yields
\begin{align}
 \lim_{\lambda\to\infty} P^2_{\textrm{on}}(s;\lambda) = 4\sqrt3\, \y^2 s\, e^{-2 (\y s)^2}  \erf(\y s)\,,
\end{align}
with
\begin{align}
 \y=\lim_{\lambda\to\infty}D=\frac{2+3\sqrt{2}\,\mbox{arccot}\sqrt{2}}{2\sqrt{3\pi}} \approx 0.7510 \,.
\end{align}

For small spacings $s$ and non-zero $\lambda$, the spacing distribution $P^2_{\textrm{on}}$ is proportional to $s^2$, whereas in the limit $\lambda \to 0$ the first term in the Taylor expansion is proportional to $s^6$.  One can understand these repulsion strengths from the Vandermonde determinant in the joint probability density alone, i.e.\ by focussing on the linear repulsion of every eigenvalue pair and neglecting the exponential and erfc factors (which only contribute significantly at larger spacings). The relevant integral is
\begin{align}
& \int_{0}^{\sp} \!\!da
(z^*-\theta_2)(z^*-\theta_1)(z-\theta_2)(z-\theta_1)(\theta_2-\theta_1)\Big|_{z=\theta_1+a+ib,\,\theta_2-\theta_1=S}\notag\\
& =  \sp\int_{0}^{\sp} \!\!da\, \left[(a-S)^2+b^2\right](a^2+b^2)
= b^4 \sp^2+\frac{2}{3}\,b^2 \sp^4+\frac{1}{30}\,\sp^6 \,.
\end{align}
This is indeed proportional to $\sp^2$ for small $\sp$, unless the distribution of $b$ becomes a $\delta$-function around zero, which happens for $\lambda\to0$. Then, the distribution is proportional to $\sp^6$.

\begin{figure}
 \includegraphics[width=0.5\linewidth]{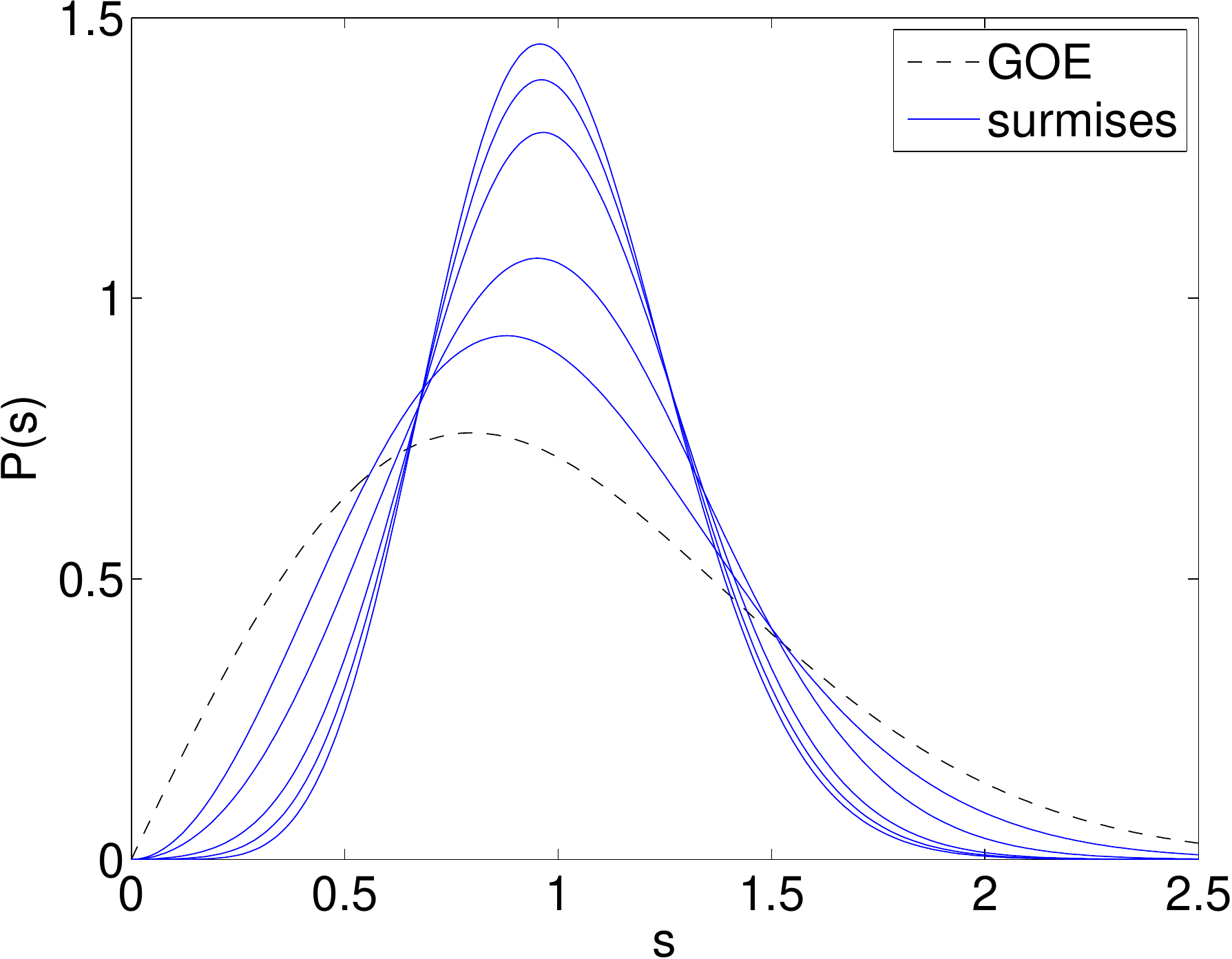}
 \includegraphics[width=0.5\linewidth]{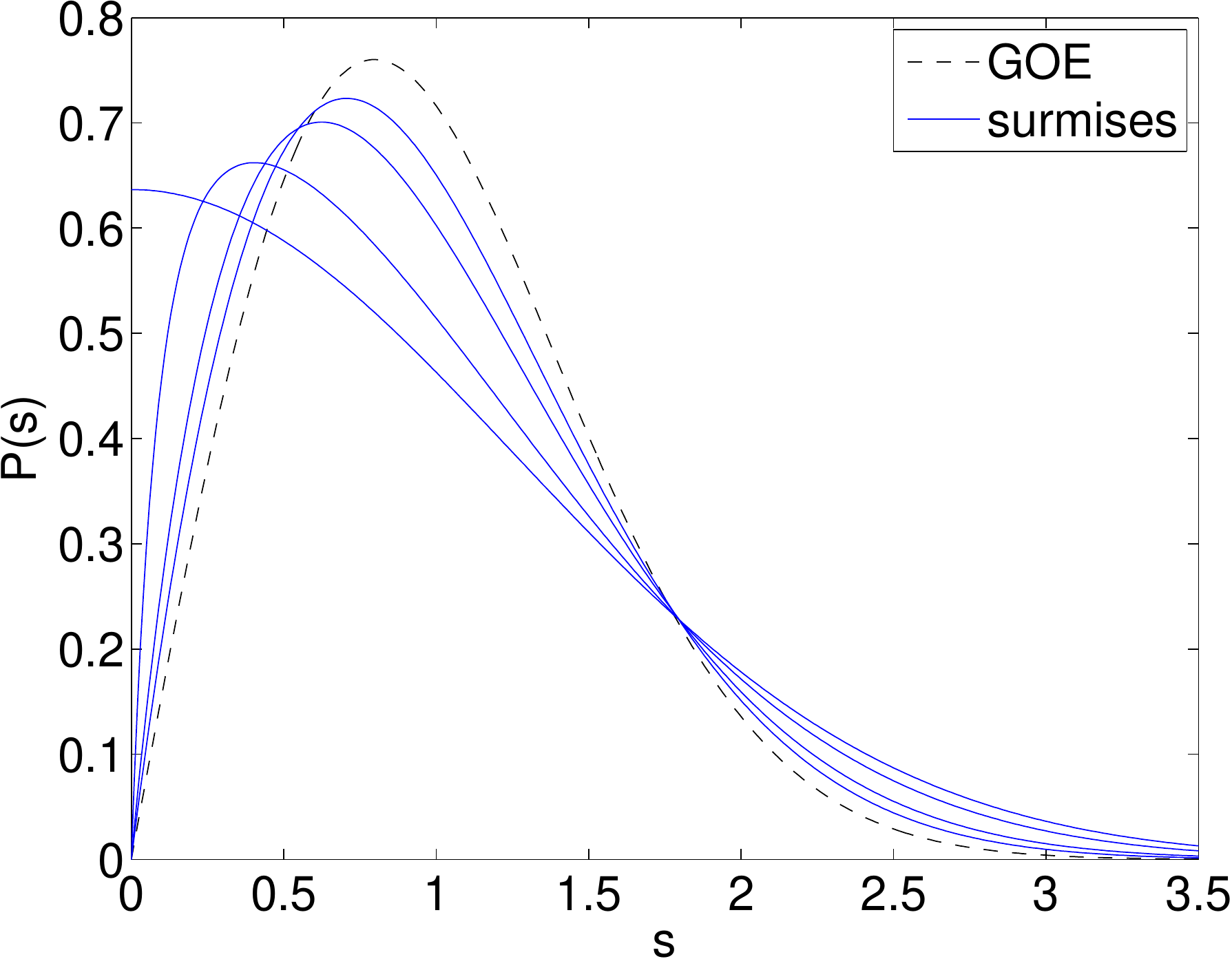}
 \caption{Left: Wigner surmise $P^2_{\textrm{on}}(s;\lambda)$, \eq{eq:surmiseon2}, for on-axis eigenvalues with an interspaced complex conjugate eigenvalue pair for $\lambda=0, 0.3, 0.5, 1, \infty$ (decreasing maxima). Right: Wigner surmise $P_{\textrm{off}}(s; \lambda)$, \eq{eq:surmiseoff}, for off-axis spacings and couplings $\lambda=0, 2, 10, \infty$ (decreasing maxima). Plotted for comparison is the Wigner surmise for the GOE (dashed), which is the surmise for the on-axis spacings with no interspaced complex pair, \eq{eq:surmiseon1}.}
 \label{fig wigner surmise}
\end{figure}

Spacings for various parameters $\lambda$ are plotted in the left panel of \fig{fig wigner surmise}.

\subsection{Off-axis}

To get a surmise for the spacings between complex conjugate eigenvalue pairs, we consider again the $2\times2$  matrix $\M$ given in \eq{eq:ortantisymm2x2}. This time, we are interested in the distribution of the imaginary spacings of this matrix. In analogy to \sect\ref{sect:onaxis1} we define $Q(\sp)$ as in \eq{eq:surmiseon_pre} with appropriate arguments of the delta and step function,
\begin{align}
 Q(\sp)&=\int_{-\infty}^\infty\!\!da\, db\, dc \,e^{-(a^2+b^2+c^2)/2}\,
 \delta\left(\sp-2\sqrt{c^2\lambda^2-a^2-b^2}\right)
 \theta\left(c^2\lambda^2-a^2-b^2\right)\,.
\end{align}
For the normalized spacing distribution we obtain with calculations similar to those of \sect\ref{sect:onaxis1},
\begin{align}\label{eq:surmiseoff}
  P_{\textrm{off}}(s; \lambda) = C\,D^2\,s\, e^{D^2 s^2}\erfc\left(Ds
\sqrt{1+\lambda^2} / \lambda\right)\,,
\end{align}
with constants
\begin{align}
 C = \frac{2}{\sqrt{1 + \lambda^2}-1} \,,\qquad
 D = C\, \frac{\lambda\sqrt{1 + \lambda^2} -
\mbox{arsinh}(\lambda)}{2\sqrt{\pi}}\,.
\end{align}
For small $s$ the level spacing is linear, just as in the GOE. This is understandable, as the derivation of $P_{\textrm{off}}$ involves an integral over one off-diagonal random number which is similar to that encountered in the derivation of the GOE spacing.  The number of degrees of freedom, which governs the level spacing at small $s$, is thus the same for both cases.

The limit for $\lambda\to0$ is
\begin{align}
 \lim_{\lambda\to 0} P_{\textrm{off}}(s; \lambda) 
  = \frac{64}{9\pi}\, s \erfc\left(\frac{4s}{3\sqrt{\pi}}\right)\,.
 \label{eq:surmise_off_limitzero} 
\end{align}
Note that in this limit the perturbation is switched off. For statistical reasons, off-axis eigenvalues will exist for arbitrary small perturbations (see \eq{lambdacrit}).  Equation \eqref{eq:surmise_off_limitzero} describes the (normalized) level spacings of these eigenvalues in the $\lambda\to 0$ limit (whereas \eq{eq:surmiseon2_smallpert} reflects the influence of these eigenvalues on the neighboring on-axis spacings in this limit). Even though the difference between \eq{eq:surmise_off_limitzero} and the GOE spacing distribution is rather small, the two are clearly distinguishable in \fig{fig wigner surmise} (right).

For $\lambda\to\infty$ the distribution $P_{\textrm{off}}$ is simply half a Gaussian,
\begin{align}
 \lim_{\lambda\to \infty} P_{\textrm{off}}(s; \lambda)
=\frac2{\pi}\,\exp\left(-\frac{s^2}{\pi}\right)\,,
\end{align}
the spacing distribution of the perturbation alone. This limit, however, is not uniform at $s=0$, since $P_{\textrm{off}}(0; \lambda)=0$ for every finite $\lambda$, whereas $\lim_{s\to 0}\lim_{\lambda\to \infty} P_{\textrm{off}}(s; \lambda)=2/\pi$. Spacings for various couplings $\lambda$ are plotted in \fig{fig wigner surmise} (right) and this discontinuity is clearly visible. A similar effect has been observed for mixed symmetry classes of (small and large) random matrices in Ref.~\cite{Schierenberg:2012ut}. There, even a Gibbs-like overshoot of the curves near $s=0$ was observed, which is absent here.

We have checked that the surmises $P^1_{\textrm{on}}$ and $P_{\textrm{off}}$ could equally well be obtained from the joint eigenvalue probability distribution of real $2\times 2$ matrices \cite{Akemann:2010mt}, i.e.\ analogues of \eq{eqn jpd}, with either two real or a pair of complex conjugate eigenvalues.

With these surmises at hand, our aim will now be to describe the level spacing distributions of large random matrices, matching the parameter $\lambda$ to the corresponding coupling parameter and to QCD spectra, where the matching will be to the chemical potential.

\section{Comparison of the surmises to large RMT spectra}
\label{sect:rmt}

To check the validity of the surmises calculated in the previous section, we applied them to the spectra of large dimensional random matrices of the form
\begin{align}\label{eq:largermtmatrix}
 \M = \S + \frac\Lambda{\sqrt{2N/\pi}}\ \A\,,
\end{align}
where $\S$ is real symmetric and taken from the GOE with matrix size $N$, whereas $\A$ is real antisymmetric with probability density
\begin{align}
 w(\A) \sim e^{-\frac12 {\rm tr}\left(\A \A^T\right)}\,,
\end{align}
i.e. all elements of $\A$ are independently Gaussian distributed with the same variance as the off-diagonal entries of $\S$ (as was also used for the small matrices in the surmises, cf.\ e.g.\ \eq{eq:probdistA}).  The coupling parameter $\Lambda$ is divided by $\sqrt{2N/\pi}$ in order to make it comparable to the one used in the surmises, $\lambda$. As far as the spacing distribution is concerned, this normalization makes $\Lambda$ a universal, $N$-independent coupling parameter, see \cite{Schierenberg:2012ut} for a detailed discussion. As also argued there, a constant spectral density is necessary to compare spacing distributions of large matrices from mixed universality classes to the corresponding distributions of small matrices.  Otherwise, different coupling strengths are mixed.  Therefore, we only evaluated eigenvalues with real part in the interval $(-\sqrt{N}/4, \sqrt{N}/4)$, i.e. around the center of the real spectrum of $\M$.  As the spectral density is not exactly constant, we measured the on-axis spacings in units of the local mean spacing (obtained by an ensemble average), which is equivalent to unfolding the spectrum.  For the off-axis spacings, no unfolding was done.

\begin{figure}[t]
 \includegraphics[width=\linewidth]{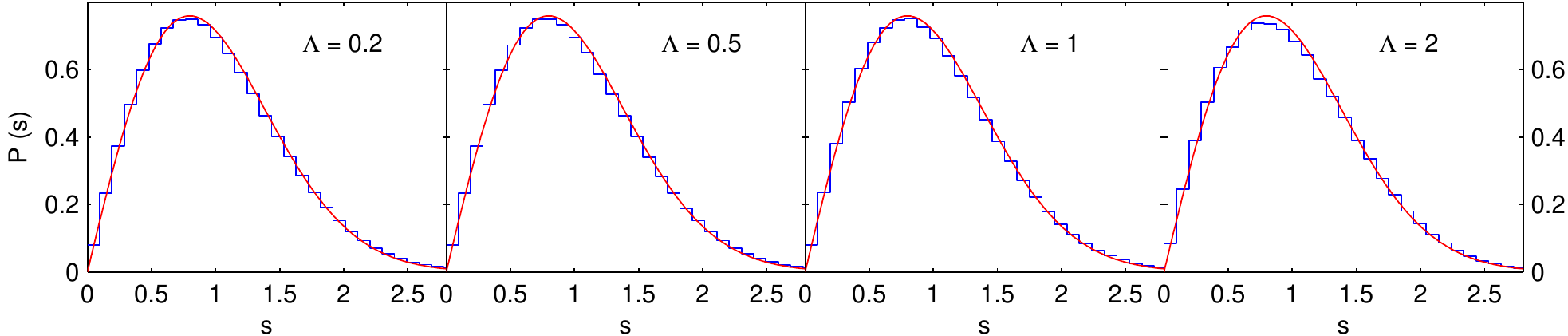}
 \includegraphics[width=\linewidth]{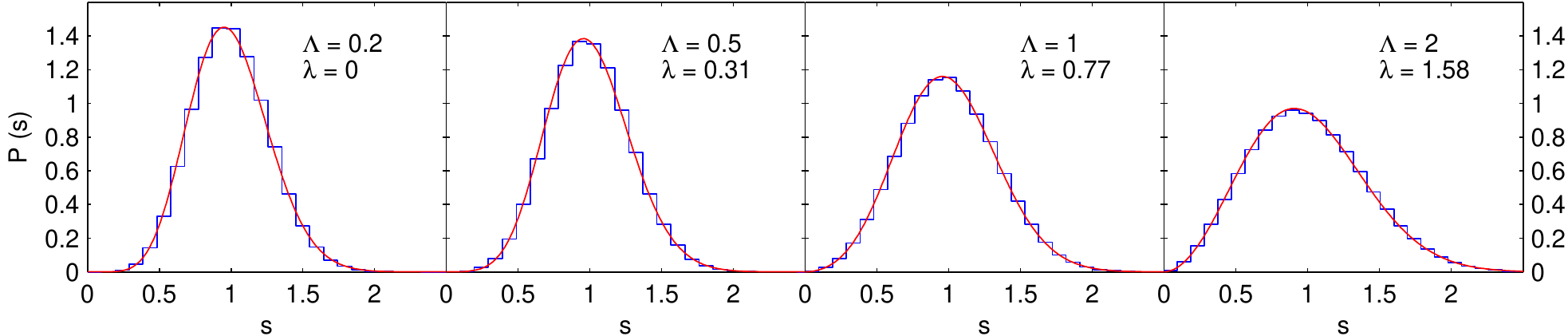}
 \includegraphics[width=\linewidth]{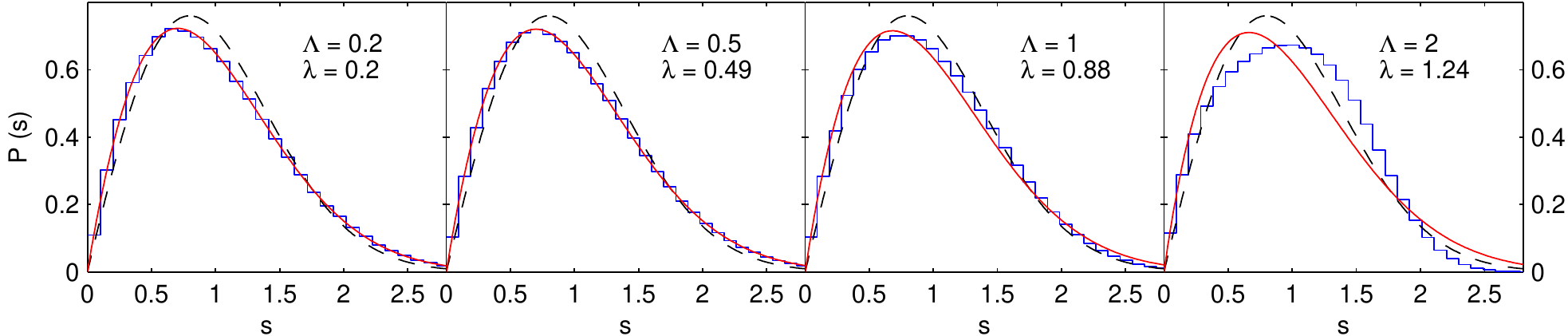}
 \caption{Spacing distributions of $400\times 400$ random matrices of the form
of \eq{eq:largermtmatrix}, with various values of the coupling parameter $\Lambda$. 
Top: on-axis spacings of type 1, surmise given by GOE,  \eq{eq:surmiseon1}. 
Middle: on-axis spacings of type 2, surmise given by \eq{eq:surmiseon2}.
Bottom: off-axis spacings, surmise given by \eq{eq:surmiseoff}; the dashed curve is the GOE spacing for comparison.  For each value of $\Lambda$, $2\cdot 10^5$ random matrices were diagonalized.}
 \label{fig:largermt}
\end{figure}

The numerically obtained spacing distributions of $\M$ are shown in \fig{fig:largermt} for various values of the coupling parameter $\Lambda$.  As can be seen, the surmises for the on-axis spacings of both types describe the data very well for coupling parameters up to $\Lambda=2$.  The coupling parameters $\lambda$ for the on-axis spacings of type 2 were obtained by a fit with least square minimization.

For the off-axis spacings we are able to predict a $2\times 2$ coupling parameter $\lambda$ through the frequency $\Pim$ of imaginary spacings: we measure the latter for the large matrices and choose $\lambda$ such that the same $\Pim$ is realized for $2\times 2$-matrices, i.e.\ we invert \eq{eq:imagspacprob} to obtain
$\lambda = \sqrt{1/(1-\Pim)^2-1}$.
The surmises for the off-axis spacings obtained this way match the numerical data very well for $\Lambda=0.2$ and $0.5$, differ slightly for $\Lambda=1$ and are far off for $\Lambda=2$. Although the coupling parameter could also be determined by a fit to the level spacing data, we observed that this did not yield an improved estimate. Note that the surmises' maxima are always left of the maximum of the GOE, cf.\ \fig{fig wigner surmise} right, in contrast to the $\Lambda=2$  maximum. Not surprisingly, the surmise for the off-axis spacing does not work for large coupling parameters, i.e.\ for large perturbations. The reason for the earlier break down of the off-axis surmise could be related to the additional degree of freedom of the eigenvalues when they move in the complex plane.

\section{Application to two color QCD with chemical potential}
\label{sect:qcd}

\subsection{Symmetry and hermiticity of the Dirac operator}
\label{subsect:symm_and_herm}

As far as anti-unitary symmetries are concerned, the surmises discussed in the previous sections should approximate the spacing distributions for two-color QCD with a real quark chemical potential, as will be analyzed now. 

\subsubsection{Continuum}
\label{sect:symmcont}

In the presence of a chemical potential, the Euclidean-space Dirac operator in continuum two-color QCD is given by
\begin{align}
D = \gamma_\nu D_\nu + m + \mu \gamma_4\,,
\end{align}
with covariant derivative
\begin{align}
D_\nu = \partial_\nu + i A^a_\nu(x) \tau_a\,,
\end{align}
where $\tau_a$ are the Pauli matrices, the generators of \sutwo. The massless Dirac operator at zero chemical potential is an anti-hermitian operator, which has purely imaginary eigenvalues. The mass merely shifts the spectrum, while the chemical potential distorts it nontrivially. Since the former is irrelevant for level spacings, we will set $m=0$ for the following discussion.

The anti-unitary symmetry of the \sutwo{} Dirac operator is based on the pseudo-real nature of \sutwo{} implying
\begin{equation}
 (i\tau_a)^\ast=\tau_2^\dagger(i \tau_a)\tau_2\,,
\label{eqn tau symm} 
\end{equation}
as well as on the charge conjugation properties of gamma matrices
\begin{equation}
 (i\gamma_\nu)^\ast=C^\dagger(i \gamma_\nu) C\qquad \mbox{with}\qquad
 C= \gamma_2\gamma_4\,,
\label{eqn charge conj}
\end{equation}
where we use the Weyl representation
\begin{equation}
 \gamma_\nu=\left(
 \begin{array}{cc}
  0 & (i\tau_i,1_2) \\ (-i\tau_i,1_2) & 0
 \end{array}
 \right)\,.
\end{equation}

Making contact to the real random matrices considered so far is easiest after multiplication with the imaginary unit,
\begin{equation}
 iD=i\gamma_\nu\otimes\big(\partial_\nu 1_2+A_\nu^a(x) i\tau_a\big)
 +i\mu\gamma_4\otimes 1_2\,,
\label{eqn iD}
\end{equation}
where we have separated the spin and gauge parts explicitly. From \eq{eqn tau symm} and \eq{eqn charge conj} it follows that $iD$ obeys the GOE symmetry
\begin{equation}
 (iD)^\ast=U^\dagger (iD) U\,,\qquad U=C\otimes\tau_2\qquad  \text{with} \quad U U^\ast=1\,.
\label{eqn iD symm}
\end{equation}
As a consequence, $iD$ can be made real by a unitary transformation. We denote such an equivalence by $\sim$ and write
\begin{equation}
 iD\sim D_s+D_a\,,
\end{equation}
where we have split the operator into its real symmetric part $D_s$ and real antisymmetric part $D_a$. From the hermiticity at $\mu=0$ one identifies $iD(\mu=0)\sim D_s$ and further $i\mu\gamma_4\sim D_a$. Thus up to the factor $i$ -- which only rotates eigenvalues, but does not influence spacings -- the situation for the continuum Dirac operator is the following: for vanishing chemical potential it is real symmetric with on-axis eigenvalues obeying the GOE symmetry, whereas the chemical potential $\mu$ introduces a real antisymmetric perturbation with possible creation of complex conjugate eigenvalues. This agrees with the setting for the random matrices with $\mu$ playing the role of the matrix coupling parameters $\lambda$ and $\Lambda$.

The massless Dirac operator also satisfies chiral symmetry, $\{D,\gamma_5\}=0$, resulting in all nonzero eigenvalues coming in pairs $\pm i\ev$. In the numerical results below we will restrict ourselves  to the half of eigenvalues with positive imaginary part and consider spacings in the bulk, away from $\ev=0$. For the spacing distribution in the bulk the chiral symmetry is known to be irrelevant \cite{Fox:1964, Nagao:1991, Nagao:1992,Halasz:1995vd,Pullirsch:1998ke}.\footnote{\label{ftnote}Note that chiral symmetry implies another anti-unitarity relation, $D^\ast=(\gamma_5 U)^\dagger D(\gamma_5 U)$, which at first sight would mean that $D$ belongs to the class of real antisymmetric operators. This is, however, not the case, since for the classification one first has to reduce the operator to irreducible blocks. This is best done on $D^2$, in which chiral eigenvalue pairs $\pm i \ev$ become twice degenerate eigenvalues $-\ev^2$. As $D^2$ commutes with $\gamma_5$, its eigenmodes can be chosen to be of definite chirality. The two relations $(D^2)^\ast=U^\dagger D^2 U=(\gamma_5 U)^\dagger D^2 (\gamma_5 U)$ coincide on the subspaces of fixed chirality, where both are proportional to $u=\tau_2\otimes\tau_2$ with $uu^\ast=1$.  This qualifies $D^2$ and thus also $D$ for GOE.}

\subsubsection{Lattice}

When simulating the theory on a space-time lattice, the Dirac operator has to be discretized accordingly. Lattice discretizations of the Dirac operator typically alter some of its symmetry properties, such that its anti-unitarity properties can differ from those in the continuum\footnote{as is the case for the staggered Dirac operator \cite{Follana:2006zz, Bruckmann:2008xr}} and have to be analyzed carefully. 

To preserve the chiral symmetry of the discretized Dirac operator one uses the overlap fermion formulation \cite{Narayanan:1993sk,Narayanan:1994gw,Neuberger:1997fp}, whose definition, in the presence of a quark chemical potential $\mu$, is given by \cite{Bloch:2006cd}
\begin{align}
\Dov(\mu) = 
1 + \gamma_5\sgn(\gamma_5\Dw(\mu))\,,
\label{Dovmu} 
\end{align}
where we have put the lattice spacing to unity, $a=1$, $\sgn$ is the matrix sign function satisfying $(\sgn A)^2=1$, and
\begin{align}
\Dw(\mu)
= \, & 1 - \kappa \sum_{i=1}^3 \left( T_i^+ + T_i^-\right) 
- \kappa \left( {e^{\,\mu}}\, T_4^+ + {e^{-\mu}}\, T_4^-\right)
\label{DW}
\intertext{with}
&(T^{\pm}_\nu)_{yx} = (1 \pm
\gamma_\nu) \, U_{x,\pm\nu} \, \delta_{y,x\pm\hat\nu}
\end{align}
is the Wilson Dirac operator at non-zero chemical potential \cite{Hasenfratz:1983ba} with hopping parameter $\kappa = 1/(8+2m_\text{w})$, Wilson mass $m_\text{w} \in (-2,0)$ and gauge configurations $U_{x,\pm\nu}\in$
\sutwo, where $U_{x,-\nu} \equiv {U^\dagger}_{\!\!\!{x-\hat\nu,+\nu}}$.  The exponential factors $e^{\pm\mu}$ implement the quark chemical potential on the lattice. For $\mu = 0$ the argument of the sign function in \eq{Dovmu} is hermitian, while for $\mu\ne 0$ it is non-hermitian. 

To actually compute the overlap operator we need to define the matrix sign function for a general complex matrix $A$ of dimension $n$. A generic matrix function $f(A)$ can be defined by
\begin{align}
    \label{eq:matfun}
    f(A) &= \frac{1}{2\pi i} \oint_\Gamma f(z)(zI-A)^{-1}dz\,,
\end{align}
where $\Gamma$ is a collection of contours in $\mathbb{C}$ such that $f$ is analytic inside and on $\Gamma$ and such that $\Gamma$ encloses the spectrum of $A$. If $A$ is diagonalizable, i.e., $A=U\Lambda U^{-1}$, with diagonal eigenvalue matrix $\Lambda = \diag(\lambda_1, \ldots, \lambda_n)$ and $U\in \Gl(n,\mathbb{C})$, then this general definition can be simplified to the well-known spectral form
\begin{align}
    f(A) &= Uf(\Lambda)U^{-1}\,,
\label{fAdef}
\end{align}
with
\begin{align}
    f(\Lambda) &= \diag\left(f(\lambda_1),\ldots,f(\lambda_n)\right)\,.
\end{align}
If $A$ cannot be diagonalized, a spectral definition of $f(A)$ can still be derived using the Jordan decomposition \cite{Golub}.
For hermitian $A$ the eigenvalues are real and their sign is defined by $\sgn(x)=\pm 1$ for $x\gtrless 0$ with $x \in \mathbb{R}$, such that \eq{fAdef} readily defines the matrix sign function.
For non-hermitian $A$ the eigenvalues are complex and require a definition of $\sgn(z)$ for $z \in \mathbb{C}$. The sign function needs to satisfy $(\sgn z)^2=1$ and reproduce the usual $\sgn(x)$ for real $x$. We define \cite{Bloch:2007xi}
\begin{align}
    \sgn(z) &= \frac{z}{\sqrt{z^2}} = \sgn\left(\re z\right)\,,
    \label{eq:sgnz}
\end{align}
where the cut of the square root is chosen along the negative real axis. This choice, although not unique, gives the correct physical result for the overlap Dirac operator in \eq{Dovmu} (see Ref.~\cite{Bloch:2007xi}).

To investigate the anti-unitarity properties of the overlap operator we first observe that links $U_{x,\pm\nu}$ as elements of the \sutwo{} group obey
\begin{equation}
 U_{x,\pm\nu}^\ast=\tau_2^\dagger\, U_{x,\pm\nu}\, \tau_2
\end{equation}
(as they can be written in terms of $1_2$ and $i\tau_a$). For the Wilson Dirac operator multiplied by $\gamma_5$ one can show that
\begin{align}\label{eq:inv ordering}
\hspace{4cm} (\gamma_5\Dw)^\ast = U^\dagger (\Dw\gamma_5) U\qquad\mbox{(note the inverted ordering)}
\end{align}
with $U$ as in the continuum, see \eq{eqn iD symm}. 

For the overlap operator we have to defer the discussion of its anti-unitary symmetries, because it has no definite hermiticity, even at zero chemical potential. Instead it fulfills the Ginsparg-Wilson relation \cite{Ginsparg:1981bj},
\begin{align}
  \{\Dov,\gamma_5\}=\Dov\gamma_5\Dov\,, 
\end{align}
which, together with the $\gamma_5$-her\-mi\-ti\-city, $\gamma_5\Dov=\Dov^\dagger\gamma_5$, valid at vanishing $\mu$, forces the eigenvalues on a circle $\ev=1+\exp(i\varphi)$ in the complex plane. 
When investigating the spectrum of the overlap operator, we will rather consider a related operator \cite{Neuberger:1997bg}
\begin{equation}\label{eq:overlapproj}
 D_p=\frac{2\Dov}{2-\Dov}=2\,
 \frac{1+ \gamma_5\sgn(\gamma_5\Dw)}{1- \gamma_5\sgn(\gamma_5\Dw)}\,,
\end{equation}
which satisfies exact chiral symmetry, $\{D_p,\gamma_5\}=0$, at the price of being non-local. For $\mu=0$ this operator is anti-hermitian and projects the eigenvalues of $\Dov$ from the circle onto the imaginary axis. 
At non-zero chemical potential $D_p$ loses its anti-hermiticity and the eigenvalues come in three types: pairs of opposite real eigenvalues, pairs of complex conjugate imaginary eigenvalues and quartets of complex conjugate/opposite eigenvalues $(\lambda,-\lambda,\lambda^\ast,-\lambda^\ast)$.
To form the complex quartets at finite $\mu$, typically two pairs of purely imaginary eigenvalues need to come together to allow for its creation;  in our RMT discussion the former were called \textit{on-axis} and the latter \textit{off-axis}.

The anti-unitarity symmetry of $D_p$ follows from that of $\gamma_5\Dw$, \eq{eq:inv ordering}, and the sign function thereof (and the facts that $U$ commutes with $\gamma_5$ and that $\sgn$ and $\gamma_5$ are their own inverses):
\begin{align}
 \gamma_5\sgn^\ast(\gamma_5\Dw) &= U^\dagger \gamma_5\sgn(\Dw\gamma_5) U
  =U^\dagger \sgn(\gamma_5\Dw)\gamma_5 U \,, \\
 D_p^\ast &= 2\, U^\dagger \,\frac{1+\sgn(\gamma_5\Dw)\gamma_5}
  {1-\sgn(\gamma_5\Dw)\gamma_5}\,U
  =2\, U^\dagger \,\frac{\gamma_5\sgn(\gamma_5\Dw)+1}{\gamma_5\sgn(\gamma_5\Dw)-1}\,U \notag\\
  &=- U^\dagger D_p \, U\,.
\label{eqn Dp symm}
\end{align}
Hence $i D_p$ shares the anti-unitary symmetry from the continuum, compare \eqref{eqn Dp symm} to \eqref{eqn iD symm}, and therefore GOE spacings are expected at vanishing $\mu$. Again a chemical potential destroys the hermiticity, but keeps the anti-unitary symmetry, exactly as for the continuum Dirac operator.  Note that $\mu$ enters the hermiticity breaking part of the operator $D_p$ non-linearly -- unlike $\lambda$ and $\Lambda$ in the random matrices.

\begin{table}[b]
 \begin{center}
 \begin{tabular}{c||c|c|c|c}
     $\mu$   & $0.05$  & 0.10 & 0.20 & 0.30 \\
     \hline
     \# configs & 60000  & 30000 & 20000 & 20000  
 \end{tabular}
 \caption{Values of the chemical potential and corresponding number of quenched configurations used to determine the spectral properties of the overlap operator.}
 \label{tab:sim-param}
 \end{center}
\end{table}

\subsection{Numerical results}

The quenched lattice simulations were performed on an $8^4$ lattice with the Wilson gauge action, using the USQCD lattice QCD software Chroma and QDP++ \cite{usqcd}. The software was adapted with an implementation of the overlap operator at non-zero chemical potential, based on the nested Krylov-Ritz approximation \cite{Bloch:2009uc}. We generated 60000 quenched configurations at $\beta=2.2$ using a heatbath/overrelaxation algorithm. On subsets of them we computed $\mathcal{O}(30)$ of the lowest lying eigenvalues of the overlap operator \eqref{Dovmu} with ARPACK \cite{arpack} for various values of the chemical potential and Wilson mass $m_\text{w}=-1.4$. The number of configurations used for the different chemical potentials are summarized in Table~\ref{tab:sim-param}. An animated plot of the evolution of the eigenvalues of $D_p$ with increasing chemical potential for a typical configuration can be found as ancillary material to this arXiv submission \cite{anim}.

\begin{figure}[t]
 \includegraphics[width=\linewidth]{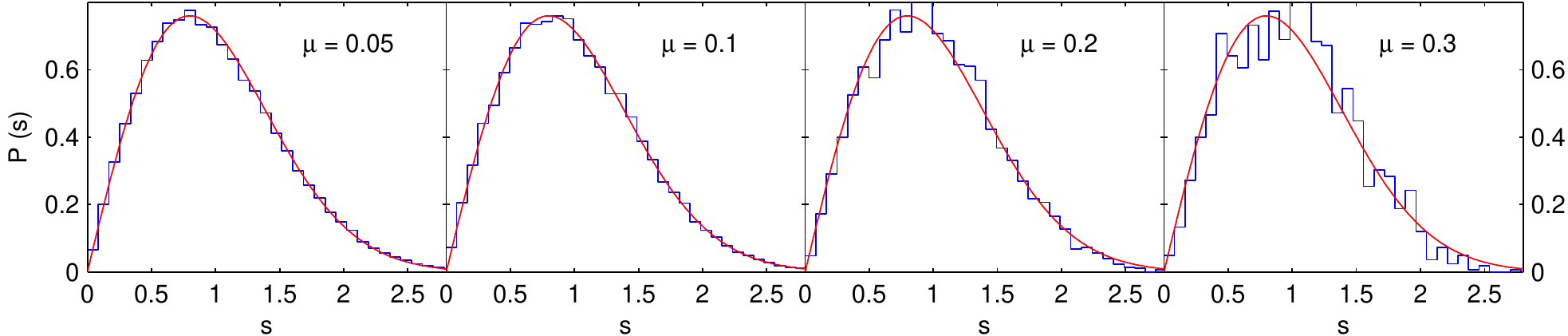}
 \includegraphics[width=\linewidth]{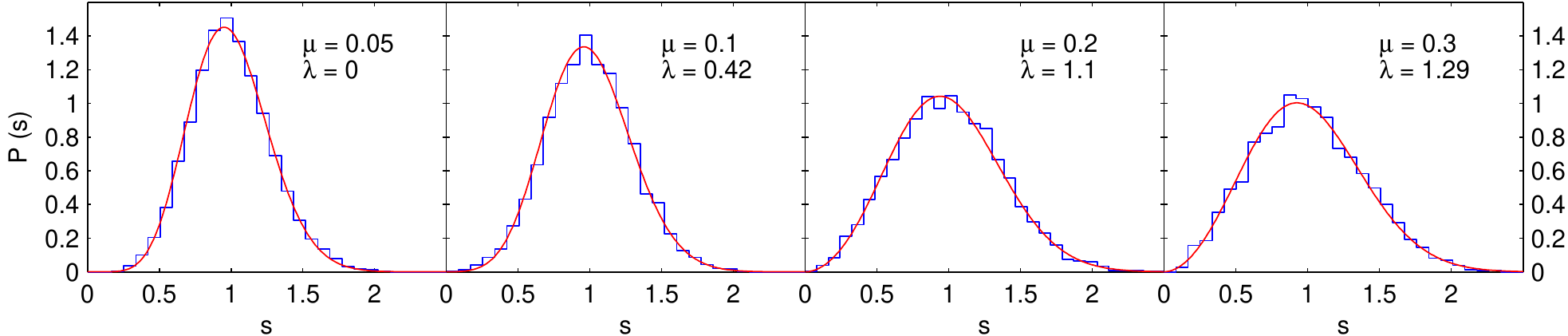}
 \includegraphics[width=\linewidth]{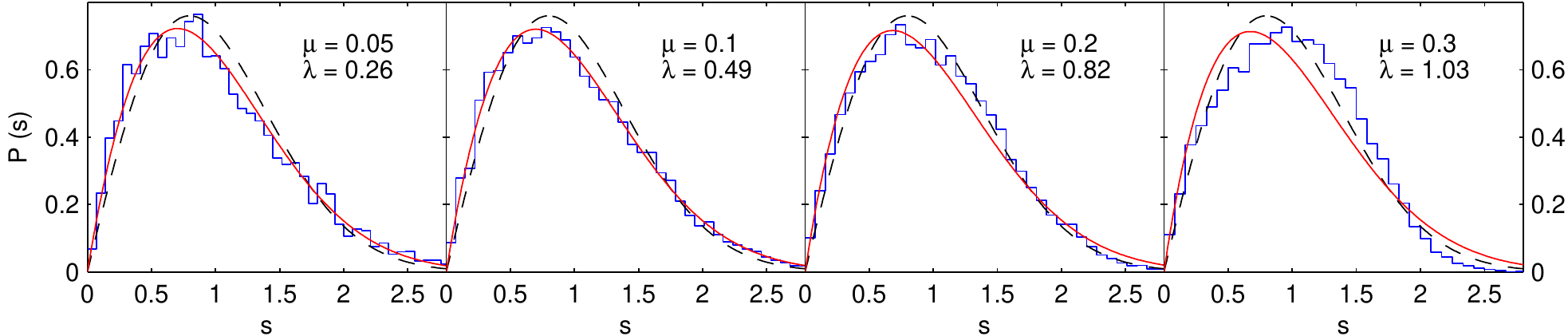}
 \caption{Spacing distributions of the overlap operator with various real values of the chemical potential $\mu$. Top: on-axis spacings of type 1, surmise given by GOE,  \eq{eq:surmiseon1}. Middle: on-axis spacings of type 2, surmise given by \eq{eq:surmiseon2}. Bottom: off-axis spacings, surmise given by \eq{eq:surmiseoff}; the dashed curve is the GOE spacing for comparison.}
 \label{fig:overlaprealmu}
\end{figure}

We measured the spacing distributions of the projected overlap operator $D_p$, defined in \eq{eq:overlapproj}, for various values of the chemical potential $\mu$.  As before, we distinguish between off-axis spacings and on-axis spacings with and without an interspaced complex pair.  We only considered spacings between eigenvalues with an imaginary part in the spectral window $(0.5, 0.6)$. This ensures that the eigenvalues stay in the bulk of the spectrum (which of course is limited on a finite lattice) and have a roughly constant eigenvalue density which is necessary to apply the surmises, as argued in \sect\ref{sect:rmt}.  As for the large random matrices, unfolding was only done for the on-axis spacings.

The results are shown in \fig{fig:overlaprealmu}. The numerical data are well matched by the surmises derived from small matrices, like for the large random matrices considered in \sect\ref{sect:rmt}. Again, the coupling parameter $\lambda$ is obtained by a fit for the on-axis spacings of type 2 and from the frequency $\Pim$ of imaginary spacings for the off-axis spacings.  For the off-axis spacings at large $\mu$ we observe that the discrepancy between the data and the surmise has the same tendency as for the large random matrices, cf.\ \fig{fig:largermt}, lower right panel. This indicates that the RMT results for large matrices are able to describe the QCD results reasonably well, even at larger coupling.  

\subsection{Imaginary chemical potential}

\begin{figure}
 \includegraphics[width=\linewidth]{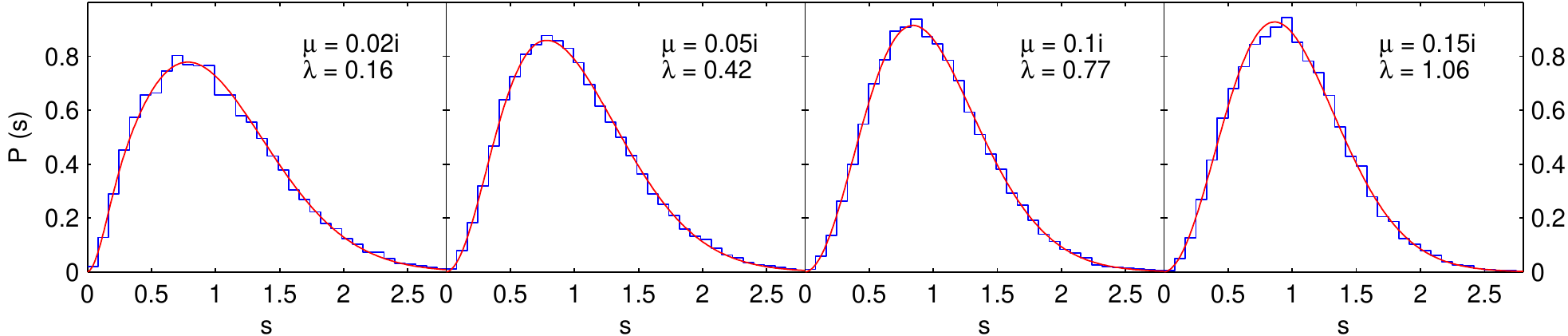}
 \caption{Spacing distributions of the overlap operator with various imaginary values of the chemical potential $\mu$ fitted by the surmise \eqref{eq:surmiseimag}. For each $\mu$, 5000 configurations have been analyzed in the  spectral window with imaginary parts of the eigenvalues between $0.65$ and $0.8$.}
 \label{fig:overlapimagmu}
\end{figure}

The introduction of an \textit{imaginary} chemical potential does not change the anti-hermiticity of the continuum Dirac operator $D$ nor of the lattice operator $D_p$. 
However, the operators $iD$ and $iD_p$ no longer obey an anti-unitary symmetry.  The random matrix ensemble appropriate for operators without anti-unitary symmetry is the GUE. Hence, with increasing imaginary chemical potential one expects a transition from GOE to GUE. Surmises for such mixed random matrices, again additive, $H_\text{GOE}+\lambda H_\text{GUE}$, have been worked out in Refs.~\cite{Lenz:1991, Schierenberg:2012ut}, with the result
\begin{equation}
 P_{\text{GOE}\to\text{GUE}}(s)= C s\, e^{-D^2s^2} \erf\left(\frac{Ds}{\lambda}\right)\,,
\label{eq:surmiseimag}
\end{equation}
where
\begin{align}
	C = 2\sqrt{1+\lambda^2}\,D^2 \,,\qquad 
  D = \frac{\sqrt{1+\lambda^2}}{\sqrt{\pi}}
  \left(\frac{\lambda}{1+\lambda^2} + \text{arccot}\lambda\right)
  \,.
\label{eq:surmiseimag_const}
\end{align}
In \fig{fig:overlapimagmu}, these distributions are compared to those of the QCD Dirac operator $D_p$ with imaginary chemical potential, were unfolding was again done by measuring the spacings in units of the local mean spacing (obtained by an ensemble average). The coupling parameter was obtained by a fit with least square minimization. The surmise agrees very well with the numerical data.

\section{Summary}
\label{sect:summary}

We have shown that eigenvalues of real symmetric matrices are attracted under real antisymmetric perturbations and are able to merge and move into the complex plain as complex conjugate pairs.  For the emergent kinds of level spacings -- on-axis with and without interspaced eigenvalue pair and off-axis -- we have derived surmises from small random matrices.  For the on-axis spacings without interspaced pair, the surmise is simply the GOE surmise, as in the unperturbed case, while for the other two we have obtained closed formulae, which depend on a coupling parameter $\lambda$.

These mixed-hermiticity surmises provide good approximations to the spacings of large random matrices in the regime of weak non-hermiticity and for the bulk of two-color QCD with small chemical potential, as expected from the anti-unitary symmetry and hermiticity properties of the Dirac operator. For on-axis spacings with an interspaced pair the surmise parameter $\lambda$ was obtained by a fit to the level spacings, while for off-axis spacings it was predicted by matching the frequency of those spacings between surmise and data.

We have also measured the level spacings for two-color QCD with imaginary chemical potential and verified that they follow a mixed-symmetry surmise. 

In contrast to the eigenvalue densities, the level spacings of asymmetric real random matrices have not been worked out analytically. However, for weak antisymmetric perturbations the analytic surmises, which approximate the level spacings of large random matrices, can be used to describe physical systems obeying the relevant anti-unitary symmetry properties.

\section*{Acknowledgements}

The spectra of the overlap operator have been calculated on the Athene and iDataCool compute clusters, which are part of the HPC infrastructure of the University of Regensburg. We would like to thank Gernot Akemann for useful discussions. We also acknowledge DFG for financial support (SFB/TR-55 and BR 2872/4-2).

\bibliography{references}
\bibliographystyle{JHEP}

\end{document}